\documentclass[aps,twocolumn,showpacs,nofootinbib]{revtex4}
\usepackage{epsfig}
\usepackage{amsmath}
\newcommand{\be}{\begin{equation}} 
\newcommand{\ee}{\end{equation}}
\newcommand{\bea}{\begin{eqnarray}} 
\newcommand{\eea}{\end{eqnarray}}

\newcommand{\dif}{\mathrm{d}}
\newcommand{\Kn}{\mathrm{Kn}}
\newcommand{\cms}{\mathrm{cms}}

\begin{document}
\title{Dynamical freeze-out criterion in a hydrodynamical description of
Au\,+\,Au collisions at $\sqrt{s_\mathrm{NN}}=200$~GeV and Pb\,+\,Pb 
collisions at $\sqrt{s_\mathrm{NN}}=2760$~GeV.}
\author{Saeed Ahmad$^a$, Hannu Holopainen$^b$ and Pasi Huovinen$^{b,c}$}
\affiliation{$^a$STEM Division, Eastfield College, 3737 Motley Drive,
Mesquite, TX 75150, USA}
\affiliation{$^b$Frankfurt Institute for Advanced Studies, Ruth-Moufang-Str.\ 1,
D-60438 Frankfurt am Main, Germany}
\affiliation{$^c$Institute of Theoretical Physics, University of Wroclaw,
pl.~M.~Borna 9, PL-50204 Wroc\l aw, Poland}

\date{\today}
\begin{abstract}
In hydrodynamical modeling of ultrarelativistic heavy-ion collisions,
the freeze-out is typically assumed to take place at a surface of
constant temperature or energy density. A more physical approach is
to assume that freeze-out takes place at a surface of constant Knudsen
number. We evaluate the Knudsen number as a ratio of the expansion
rate of the system to the pion scattering rate, and apply the constant
Knudsen number freeze-out criterion to ideal hydrodynamical
description of heavy-ion collisions at RHIC
($\sqrt{s_\mathrm{NN}}=200$~GeV) and the LHC
($\sqrt{s_\mathrm{NN}}=2760$~GeV) energies. We see that once the
numerical values of freeze-out temperature and freeze-out Knudsen
number are chosen to produce similar $p_T$ distributions, the elliptic
and triangular anisotropies are similar too, in both event-by-event
and averaged initial state calculations.
\end{abstract}

\maketitle

\section{Introduction}

The fluid-dynamical description of heavy-ion collisions at the BNL
Relativistic Heavy Ion Collider (RHIC) and the CERN Large Hadron Collider
(LHC) has been very successful in reproducing the observed particle
distributions and their anisotropies at low values of transverse
momentum~\cite{Heinz:2013th,Gale:2013da,Niemi:2014lha}. However, since
what is experimentally observed is not a particle fluid, but
individual particles, the fluid-dynamical description must break down
at some point during the evolution, the interactions must cease, and
the particles must start behaving like free-streaming particles
instead. The particles decouple from the fluid, and their momentum
distributions freeze-out---a process appropriately known as decoupling
or freeze-out.

When the freeze-out happens is not described by fluid dynamics but has
to be decided using some other model or theory. Fluid dynamics is
considered to be valid when the ratio of the microscopic to
macroscopic scales of the system---its Knudsen number---is much
smaller than one. In the context of heavy-ion collisions, fluid
dynamics has traditionally been considered to be valid until either
the mean free path of particles exceeds the size of the system, or the
expansion rate exceeds the collision rate of the
particles~\cite{Heinz:1990,Kolb:2003dz}. The Knudsen number can be
defined in several ways~\cite{Niemi:2014wta}, and thus both of these
dynamical criteria are equivalent to the requirement that the Knudsen
number is less than one. The idea of using the scattering and
expansion rates as the limit for the validity of fluid dynamics, and
thus as a decoupling criterion, is an old one~\cite{Bondorf:1978kz},
but it has been used in fluid-dynamical calculations only a couple of
times~\cite{Mishustin:1983nv,Hung:1997du,Heinz:2006ur,Eskola:2007zc,
  Molnar:2014zha}. Instead, the freeze-out is assumed to take place on
a surface of constant temperature (or density). It has been argued
that since the scattering rate depends strongly on temperature
($\propto T^3$ for a constant cross section), the freeze-out is a very
fast process, and thus a constant temperature surface is a good
approximation to the constant Knudsen number
surface~\cite{Rischke:1998fq,Kolb:2003dz,Schnedermann:1994gc}. It is
worth noticing that the well-known Gamow criterion in cosmology---that
the time when interaction ceases to be effective is determined by the
condition $t_{\mathrm{int}} \le t_{\mathrm{expan}}$, where
$t_{\mathrm{int}}$ and $t_{\mathrm{expan}}$ are the relevant interaction
and expansion timescales~\cite{cosmology}---is equivalent to
freeze-out at constant Knudsen number, and leads to decoupling at
a certain temperature only because the expansion of the universe is
taken to be uniform.

It was seen in earlier studies with optical Glauber initial profiles
that, while the constant Knudsen number surface differs significantly
from the constant temperature surface, the effect on observable
particle $p_T$ distributions is small~\cite{Eskola:2007zc} and that
elliptic flow of charged hadrons shows sensitivity to the freeze-out
criterion only at large values of transverse momentum or rapidity, or
in peripheral collisions~\cite{Molnar:2014zha}. However, in
contemporary event-by-event hydrodynamical calculations, the flow
develops more violently and more unevenly than when an averaged
initial state is used~\cite{Holopainen:2010gz}. Thus it is not
obvious whether the two freeze-out conditions lead to similar particle
distributions when the initial density fluctuates
event-by-event. Furthermore, the evaluation of the Knudsen number in
Refs.~\cite{Eskola:2007zc,Molnar:2014zha} was based either on
pion-pion scattering ignoring all other scattering processes and the
chemical nonequilibrium during the hadronic
stage~\cite{Eskola:2007zc}, or on assumed temperature dependence of
the shear viscosity coefficient~\cite{Molnar:2014zha}. Thus it is
unknown how more sophisticated calculations of the microscopic scale
would affect the results.

In this work we further study whether the freeze-out criterion has any
observable effects. We evaluate the $p_T$ differential elliptic flow
$v_2(p_T)$ of identified particles (pions and protons) in
$\sqrt{s_\mathrm{NN}}=200$~GeV Au\,+\,Au (RHIC) and 
$\sqrt{s_\mathrm{NN}}=2760$~GeV Pb\,+\,Pb collisions (LHC) using both
constant temperature and constant Knudsen number freeze-out
criteria. To test our assumption that the large gradients in
event-by-event calculations would make the system more sensitive to
the freeze-out criterion, we model the collisions at RHIC both
event-by-event and using the averaged initial state. We use the pion
scattering rate as the microscopic scale, and calculate the rate in a
chemically frozen hadron gas from scattering cross sections, including
scatterings with all hadron species. Since our aim is not a faithful
reproduction of the data, we simplify the description by using a
simple boost-invariant ideal fluid model.

Note that in this work we use the conventional Cooper-Frye description
(see Sec.~\ref{frozen}) to evaluate the particle distributions at
freeze-out. We do not address the negative contributions\footnote{For
a recent discussion see
Refs.~\cite{Oliinychenko:2014tqa,Oliinychenko:2014ava}.}, but our
approach differs from the conventional freeze-out procedure only by
the choice of the decoupling surface.

To some extent the freeze-out problem has been solved in so-called
hybrid models, where the late stage of the evolution is described
using a Boltzmann transport
model~\cite{Hirano:2012kj,Petersen:2014yqa}. Nevertheless, the results
in these models depend on when the switch from fluid to cascade is
made~\cite{Hirano:2012kj,Huovinen:2012is}, and therefore it is
interesting to study how different criteria for particlization surface
affect the particle distributions even in the context of hybrid
models.

\section{Dynamical freeze-out criterion and scattering rate}

To maintain kinetic equilibrium in an expanding system the scattering
rate must be much larger than the expansion rate. We express this
condition as
\begin{equation}
\Kn = \frac{\theta}{\Gamma} \ll 1,
\end{equation}
where $\Gamma$ is the scattering rate and $\theta$ is the
hydrodynamical expansion rate. When $\Kn$ approaches one, there are
not enough collisions to maintain the kinetic equilibrium,
and the system freezes-out. Since $\Kn$ is a ratio between (an
inverse of) a macroscopic length scale and (an inverse of) a
microscopic length, it can be identified as a Knudsen number, which
should be much smaller than one for fluid dynamics to be valid. Based
on these considerations we define a dynamical freeze-out criterion as
a surface of constant Knudsen number $\Kn = \Kn_f$, where $\Kn_f\sim 1$.

Before one can apply this criterion, the scattering rate must be
known. We evaluate the pion scattering rate in hadron resonance gas
and use it in our freeze-out criterion for all particles. One could
argue that we should calculate the scattering rate individually for
each particle species and decouple them separately at the
corresponding Knudsen number. However, in order to be consistent, one
should also remove the decoupled particles from the fluid and model
the interaction between the fluid and the decoupled
particles\footnote{See the discussion about ``pion wind'' in
  Ref.~\cite{Hung:1997du}.}, which are not in equilibrium
anymore. This cannot be consistently implemented (at least not easily)
in the hydrodynamical framework and thus we make the simplifying
assumption that the whole system decouples when the most abundant
particles, \emph{i.e.} pions, do.

\subsection*{Scattering rate of pions}

Here we calculate the average scattering rate of pions in hadron
resonance gas in kinetic equilibrium. The rate is obtained
from~\cite{Tomasik:2002qt,Ftacnik:1988ar,Prakash:1993bt,Zhang:1998tj}
\begin{equation}
\label{rate1}
\begin{split}
\Gamma = \frac{1}{n_\pi(T,\mu_\pi)} \sum_i& \int \dif^3 p_\pi \dif^3 p_i\,
f_\pi(T,\mu_\pi) f_i(T,\mu_i) \\
& \times\frac{\sqrt{(s-s_a)(s-s_b)}}{2 E_\pi E_i}\,\sigma_{\pi i} (s),
\end{split}
\end{equation}
where $n_\pi$ is the density of pions, $f_{\pi/i}(T,\mu_{\pi/i})$ is
the thermal distribution function of pions/particle $i$, $T$ being
temperature and $\mu_{\pi/i}$ the chemical potential of pion/particle
$i$. $\sqrt{(s-s_a)(s-s_b)}/(2 E_\pi E_i)$ is the relative velocity
when $s$ is the square of the center-of-mass energy, 
$s_a = (m_\pi+m_i)^2$, $s_b = (m_\pi-m_i)^2$, and $E_{\pi/i}$ and
$m_{\pi/i}$ are the energy and mass of pion/particle $i$, respectively.
The pion-particle $i$ scattering cross section is labeled 
$\sigma_{\pi i}$ and the sum runs over all particle species included 
in the equation of state (EoS)~\cite{Huovinen:2009yb}.

One can perform most of the integrals analytically and, after some
algebra (see Appendix~\ref{algebra}), one arrives at:
\begin{widetext}
\begin{equation}
\label{rate}
\begin{split}
\Gamma = \frac{T}{n_\pi(T, \mu_\pi)}& \sum_i \frac{g_i}{32\pi^4}
\sum_{k=1}^\infty e^{k\mu_\pi/T}
\sum_{n=1}^\infty \frac{(\mp1)^{n+1}}{n}e^{n \mu_i /T}\\
&\times \int_{s_a}^{\infty}\! \dif s\, \frac{\sigma_{\pi i}(s)(s-s_a)(s-s_b)}
{\sqrt{rs-(r-1)(m_i^2 - rm_\pi^2)}}\, K_1\!\left(\frac{n}{T}\sqrt{rs-(r-1)
(m_i^2 - rm_\pi^2)} \right),
\end{split}
\end{equation}
\end{widetext}
where $g_i$ is the degeneracy of particle $i$, and $r=k/n$.

Cross sections are evaluated as in the UrQMD model
\cite{Bass:1998ca,Bleicher:1999xi}. Thus the largest contribution
comes from resonance formation, which is evaluated using the
Breit-Wigner formula:
\begin{equation}
\label{cross-section}
\begin{split}
\sigma_{\pi i \rightarrow R}(s) =& \frac{2 g_R + 1}{(2 g_\pi + 1)(2 g_i + 1)}
\frac{\pi}{(p_\cms(\sqrt{s}))^2} \\
&\times \frac{ \Gamma_{R \rightarrow \pi i}(\sqrt{s})\, \Gamma_{\text{tot}}
(\sqrt{s}) }{ (m_R - \sqrt{s})^2 + \Gamma_{\text{tot}}^2 (\sqrt{s}) / 4},
\end{split}
\end{equation}
where $g_{R/\pi/i}$ are the degeneracies of the
resonance/pion/particle $i$, and $p_\cms$ is the center of mass
momentum of the scattering partners (see Appendix~\ref{pcms}).
$\Gamma_{\text{tot}}(M)$ is the full decay width obtained as a sum of
all mass-dependent partial decay widths $\Gamma_{i,j}(M)$ (see
Appendix~\ref{gamma}) given by
\begin{equation}
\label{width}
\begin{split}
\Gamma_{R \rightarrow \pi i}(M) =& \Gamma_{R}^{\pi i}\, \frac{m_R}{M}\, \left(
\frac{p_\cms(M)}{p_\cms(m_R)} \right)^{2l+1} \\
&\times \frac{1.2}{1+0.2 \Big( \frac{p_\cms(M)}{p_\cms(m_R)} \Big)^{2l}},
\end{split}
\end{equation}
where $\Gamma_{R}^{\pi i}$ is the partial decay width of the resonance
into the $\pi i$ channel at the pole, $l$ the decay angular momentum of
the exit channel, and $m_R$ the pole mass of the resonance. The pole
masses and the decay widths are taken from the Particle Data
Book~\cite{PDG2004} as implemented in the calculation of the
EoS~\cite{Huovinen:2009yb}.

In addition we assume elastic meson-meson scatterings with cross
section $\sigma_{mm} = 5$~mb and elastic $\pi\pi$ scatterings with
$\sigma_{\pi\pi} = \sigma_0 \exp(-(\sqrt{s}-m_0)^2/w)$ where
$\sigma_0=15$~mb, $m_0=0.65$~GeV, and $w=0.1$~GeV$^2$. With these choices
we are able to reproduce the measured pion-pion, pion-kaon, and
pion-nucleon scattering cross sections reasonably well.

\begin{figure}[b]
\centering
\epsfig{file=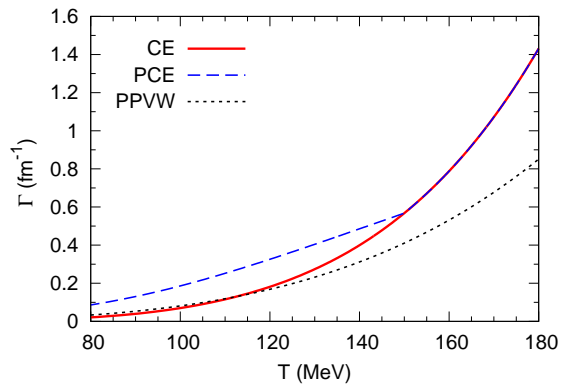, width=7.5cm}
\caption{The scattering rate of pions in both chemically equilibrated
  (CE, red solid line) and chemically frozen (PCE, blue dashed line)
  hadron resonance gas compared to the
  parametrization~\cite{Daghigh:2001gy} of the rate evaluated in
  Ref.~\cite{Prakash:1993bt} (PPVW, black dotted line).}
\label{fig:rates}
\end{figure}

In Fig.~\ref{fig:rates} we compare the evaluated scattering rates to
the rates calculated in Ref.~\cite{Prakash:1993bt}. At low
temperatures our simple approach agrees with the more sophisticated
calculation of Ref.~\cite{Prakash:1993bt}, but above $T\approx 120$
MeV temperature our rate is larger simply because we include more
states in the calculation, and thus the density of scattering partners
is larger at large temperatures. Moreover, the scattering rate in a
chemically frozen hadron gas is larger than the rate in a chemically
equilibrated hadron gas due to larger particle densities.

\section{Hydrodynamical framework}

We use an updated version of the event-by-event ideal hydrodynamical
framework developed in Ref.~\cite{Holopainen:2010gz}.

\subsection{Ideal hydrodynamics}

We solve the ideal hydrodynamical equations
\begin{equation}
\begin{split}
\partial_\mu T^{\mu\nu} = 0, \\
\partial_\mu j^\mu = 0,
\end{split}
\end{equation}
where $T^{\mu\nu} = (\epsilon + P) u^\mu u^\nu - P g^{\mu\nu}$ is the
ideal energy-momentum tensor, $j^\mu = n_B u^\mu$ the net-baryon
current, $\epsilon$ is the energy density, $P$ is the pressure,
$u^\mu$ is the fluid four-velocity and $n_B$ the net-baryon
density. We use two different equations of state (EoS): {\it (i)}
\emph{s95p}-v1, which is always in chemical equilibrium, and {\it(ii)}
\emph{s95p}-PCE-v1, which has a chemical freeze-out at temperature
$T_{\text{chem}} = 150$~MeV~\cite{Huovinen:2009yb}. Both of these EoSs
assume zero net-baryon density.

We concentrate on the midrapidity region, where boost-invariance is a
reasonable assumption at the LHC and full RHIC energies. This assumption
reduces the number of dimensions in evolution equations to 2+1. We use
the sharp and smooth transport algorithm \textsc{(shasta)}~\cite{Boris} to solve
the equations in hyperbolic coordinates, where one uses 
$\tau = \sqrt{t^2 - z^2}$ and 
$\eta_s = \frac{1}{2} \log \frac{t+z}{t-z}$ 
instead of time $t$ and longitudinal coordinate $z$. At the
antidiffusion stage of \textsc{shasta} we use DeVore limiter~\cite{DeVore},
which is a modified version of the Zalesak multidimensional 
limiter~\cite{Zalesak}.

\subsection{Freeze-out}
\label{frozen}

We employ two different freeze-out criteria. One is the conventional
constant temperature criterion, and the other the dynamical
criterion, where we assume freeze-out at constant Knudsen number
$\Kn$. The hydrodynamical expansion rate is needed to obtain the Knudsen
number, and in the boost invariant case it is calculated
as~\cite{Dumitru:1999ud}
\begin{equation}
\theta = \partial_{\mu} u^\mu = \partial_\tau u^\tau + \partial_x u^x +
\partial_y u^y + u^\tau/\tau.
\end{equation}
In both cases the freeze-out surface elements $\dif\Sigma_\mu$ are
obtained using \texttt{CORNELIUS++}
subroutine~\cite{Huovinen:2012is}. After the surface elements are
found, we calculate the thermal spectrum of hadron species
$i$ using Cooper-Frye prescription:
\begin{equation}
E\,\frac{\dif^3 N_i}{\dif^3 p} = \int_\Sigma \dif\Sigma_\mu p^\mu\, f_i(x,p),
\end{equation}
where $f_i(x,p)$ is the thermal distribution function of hadron $i$
and $p^\mu$ is the four-momentum of the hadron. At this stage we use
hadron gas EoS at nonzero net-baryon densities to convert the energy
and net-baryon density to temperature and chemical potentials. Since
the EoS during the fluid-dynamical evolution does not allow finite
net-baryon density, this procedure is not fully consistent, but the
violation of conservation laws is very small at RHIC and even smaller
at the LHC.

After the thermal distributions of all hadron species have been
evaluated, we sample individual hadrons as described in
Ref.~\cite{Holopainen:2010gz}. All strong and electromagnetic two- and
three-particle decays are then calculated, and the daughter particles
added to the respective thermal ensembles. Note that unlike in
Ref.~\cite{Holopainen:2010gz}, we no longer use \textsc{pythia} to
handle the decays, but evaluate the decays of all the resonances
included in the EoS. When evaluating the charged particle
multiplicities we sample hadrons within an interval $|y| < 3$ to make
sure that at midrapidity the system looks boost invariant after the
decays as well. However, when we consider the identified particle
$p_T$ spectra and flow coefficients, we take all particles into
account regardless of their rapidity to achieve better statistics.

\subsection{Initial state and centrality class definitions}

Initial state and centrality classes are defined using the Monte Carlo
(MC) Glauber model described in Ref.~\cite{Holopainen:2010gz}. 
Nucleons are randomly distributed to nucleus using a standard
two-parameter Woods-Saxon potential. Two nucleons from different nuclei
collide if their transverse distance $r_d < \sqrt{\sigma_{NN}}/\pi$,
where $\sigma_{NN}$ is the inelastic nucleon-nucleon cross section. We
take $\sigma_{NN} = 42$~mb at $\sqrt{s_{NN}}=200$~GeV and 
$\sigma_{NN} = 64$~mb at $\sqrt{s_{NN}}=2760$~GeV. Here we neglect
nucleon-nucleon correlations and finite-size effects since their
effects on anisotropies at mid-central collisions were found to be very
small~\cite{Alvioli:2011sk,Denicol:2014ywa}.

Multiplicity is taken to be proportional to the number of ancestors,
$N_{\text{anc}}$, which is a weighted sum of the number of
participants, $N_{\text{part}}$, and the number of binary collisions,
$N_{\text{bin}}$, defined as:
\begin{equation}
N_{\text{anc}} = (1-f) N_{\text{part}} + f N_{\text{bin}},
\end{equation}
where $f$ is the fraction of the binary collision contribution. This
fraction $f$ is chosen to reproduce the centrality dependence of
multiplicity.

In principle, when a fit to the multiplicity data is made, one should
first generate events with a certain $f$, sort the events according to
their centrality, and then calculate the average number of ancestors
in each centrality bin. Unfortunately this is a very time-consuming
procedure because a large number of events must be made for the
centrality class definitions. Thus our approach here is to fix the
centrality classes using fixed impact parameter intervals. Because the
average number of participants and binary collisions is now known at
each centrality bin, a $\chi$-squared fit can be easily made to fit
the ratio $f$. This approximation is well justified, because average
$N_{\text{part}}$ and $N_{\text{bin}}$ values are not sensitive to the
centrality class definition.

After the fraction of binary collisions, $f$, is determined, we
convert the centrality classes to number of ancestors intervals. To
fix $f$, we used the STAR Collaboration data~\cite{Abelev:2008ab} from
RHIC, and the ALICE Collaboration data~\cite{Aamodt:2010cz} from the
LHC. We neglected the most-peripheral centrality classes starting from
60\% centrality, since we do not expect hydrodynamics to be
applicable for peripheral collisions. Our result for RHIC is
$f=0.088$ and at the LHC we obtain $f=0.17$.

The initial entropy density distribution $s(x,y)$ for a single event
is taken to be
\begin{equation}
s(x,y) = \frac{K_{\text{sd}}}{\sqrt{2\pi\sigma^2}} \sum w_i
\exp \Big(- \frac{(x-x_i)^2 + (y-y_i)^2}{2\sigma^2} \Big),
\end{equation}
where the sum runs over all participants and binary collisions, $w_i$
is the weight ($(1-f)$ for participant and $f$ for binary collision),
$x_i$ and $y_i$ are the transverse coordinates of a participant or a
binary collision, and $\sigma$ is a Gaussian smearing parameter
controlling the shape of the distribution. The overall normalization
constant $K_{\text{sd}}$ is fixed to reproduce the observed
multiplicity in the 0-5\% most-central collisions. In this work we use
$\sigma=0.8$~fm. We do not study the dependence of the results on
$\sigma$ because smaller width of the Gaussians leads to a formation
of very small scale structures on the constant Knudsen number surface;
see Ref.~\cite{Holopainen:2012id}. The scale of these structures is
smaller than the mean free path of pions, and thus we do not consider
them physical. At this stage we do not consider it worth the effort to
improve the freeze-out criterion to remove these structures since
further studies should be carried out using viscous hydrodynamics, and
dissipation is known to smear small scale structures anyway.

To calculate the average initial state, we average 1000 MC Glauber
initial states. In this procedure impact parameters in each event are
aligned. We first obtain an averaged entropy density profile and then
use the EoS to convert it to energy density profile, which is the
actual initial condition for hydrodynamics.

\section{Results}

We concentrate on the effects of the freeze-out criterion on particle
distributions and their anisotropies and do not aim to faithfully
reproduce the data. We compare the calculated $p_T$ distributions to
the data to show that our parameter choices are reasonable, but do not
compare elliptic flow nor triangular flow with the data to avoid the
need to evaluate the anisotropy the same way the particular data set
was analyzed. It was seen in Ref.~\cite{Molnar:2014zha} that the
favored freeze-out temperature and Knudsen number do not depend on
centrality in the 0-50\% centrality range where fluid dynamics works
best. We do not expect event-by-event fluctuations to change this
behavior, and therefore do not study the centrality dependence of the
$p_T$ spectra or anisotropies in detail. Instead, we mostly
concentrate on the 20-30\% centrality bin, and leave the study of $p+A$
and peripheral $A+A$ collisions for a later work.

\subsection{Averaged initial state in $\sqrt{s_{NN}}=200$~GeV Au+Au collisions
at RHIC}

\begin{figure}[t]
\centering
\epsfig{file=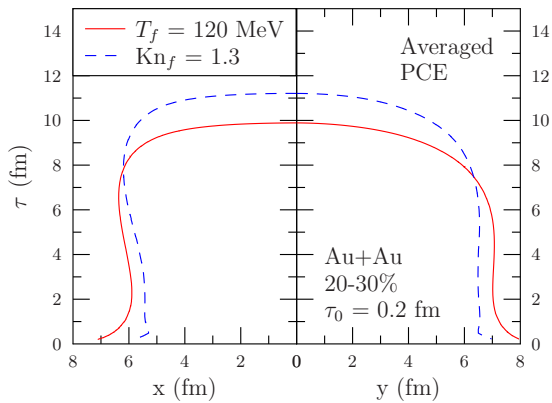, width=7.5cm}
\caption{Constant temperature (solid red curve) and constant Knudsen
number (dashed blue curve) freeze-out surfaces in
$\sqrt{s_{NN}}=200$~GeV 20-30\% central Au\,+\,Au collisions. Surfaces
are shown along the $x$ and $y$ axes. }
\label{fig: rhic surface pce}
\end{figure}

To visualize how the freeze-out surface depends on the freeze-out
criterion, we show the constant temperature and constant Knudsen
number freeze-out surfaces in Fig.~\ref{fig: rhic surface pce}. The
surfaces are calculated using an average initial state for a
$\sqrt{s_{NN}}=200$~GeV 20-30\% central Au\,+\,Au collision and chemically
frozen \emph{s95p}-PCE-v1 EoS. The constant Knudsen number surface is
closer to the center of the system, and thus the edges of the system
are hotter and the maximum flow velocity is lower than on the constant
temperature surface. On the other hand, the system lives longer, and
the center decouples at lower temperature. Similar behavior can be
seen at the LHC energy as well, and when chemical equilibrium is
assumed.

\begin{figure}[t]
\centering
\epsfig{file=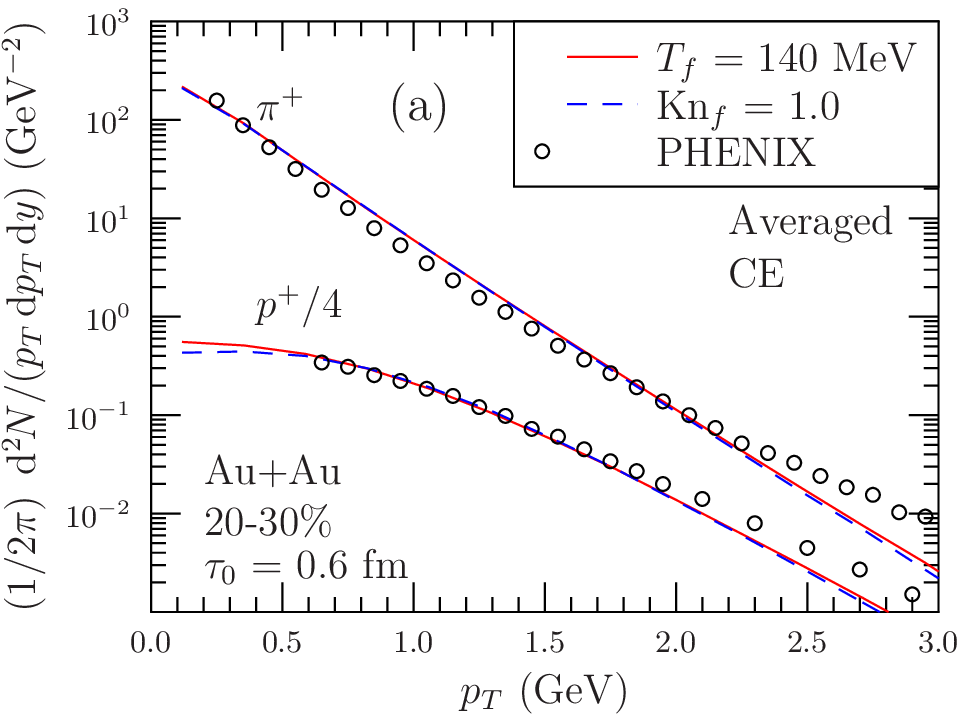, width=7.5cm}
\epsfig{file=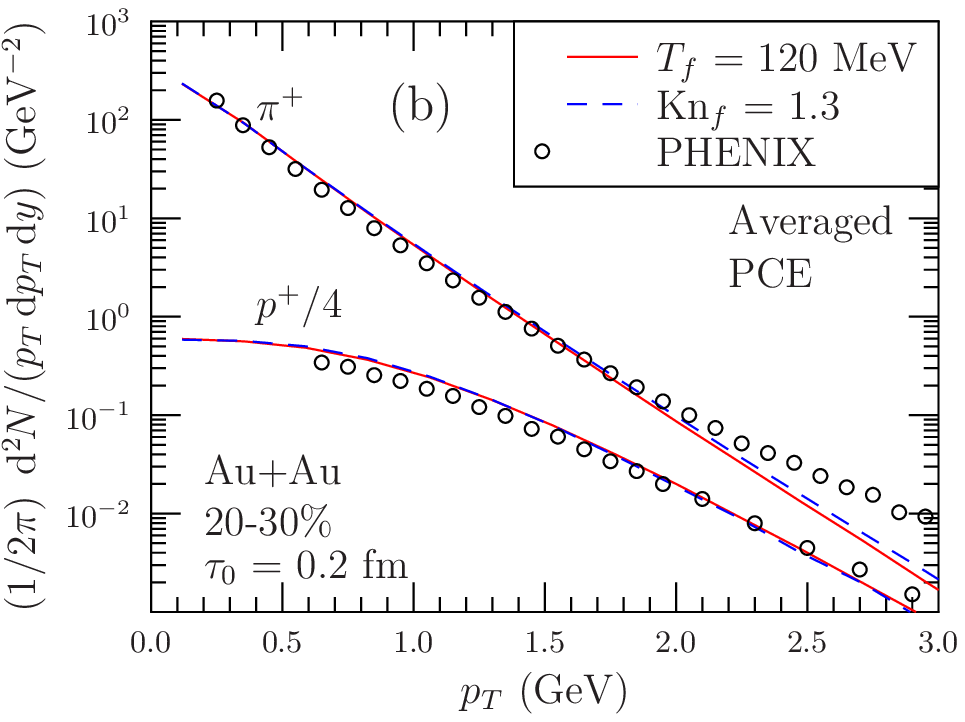, width=7.5cm}
\caption{Transverse momentum spectra of positive pions and protons
in $\sqrt{s_{NN}}=200$~GeV 20-30\% central Au\,+\,Au collisions
assuming (a) chemically equilibrated or (b) chemically frozen
(bottom) EoS. The solid red line corresponds to the results
obtained using freeze-out at constant temperature and the dashed
blue line using freeze-out at constant Knudsen number. The data
are from the PHENIX Collaboration \cite{Adler:2003cb}.}
\label{fig: rhic spectra}
\end{figure}

In Fig.~\ref{fig: rhic spectra} we show the transverse momentum
spectra of positive pions and protons in 20-30\% centrality class. The
calculations were performed either using the EoS \emph{s95p}-v1, which
assumes chemical equilibrium (Fig.~\ref{fig: rhic spectra}, top panel),
or the \emph{s95p}-PCE-v1 EoS (Fig~\ref{fig: rhic spectra}, bottom
panel), which assumes chemical freeze-out at $T_{chem}= 150$~MeV. With
\emph{s95p}-v1 EoS the initial time is the conventional 
$\tau_0 = 0.6$~fm, and the freeze-out temperature $T_f = 140$~MeV and
Knudsen number $\Kn_f = 1$ lead to almost identical pion and proton
distributions which reproduce the data reasonably well.

The assumption of separate chemical freeze-out (Fig.~\ref{fig: rhic
  spectra}, bottom) necessitates the use of an earlier initial time
$\tau_0 = 0.2$~fm to make the proton spectrum hard
enough\footnote{Later freeze-out, \emph{i.e.} lower freeze-out
  temperature or larger freeze-out Knudsen number, would make the pion
  spectrum too soft; see discussions in
  Refs.~\cite{Hirano:2005wx,Huovinen:2007xh}.}. When chemical
equilibrium has been lost, the temperature decreases faster with
decreasing energy density than in chemical equilibrium. This
necessitates the use of lower freeze-out temperature $T_f=120$~MeV,
and larger freeze-out Knudsen number $\Kn_f = 1.3$ to get sufficient
transverse flow to reproduce the data. Since $\Kn_f$ is a free
parameter of the order of one, and the assumption of chemical
equilibrium until the end of the evolution somewhat unphysical, it is
acceptable that $\Kn_f$ is different for CE and PCE EoSs.

As shown it is possible to find constant temperature and constant
Knudsen number values for freeze-out, which give similar pion and
proton spectra. This is a nontrivial result, since the corresponding
freeze-out surfaces are different. On a constant Knudsen number
surface the average flow velocity is lower, and the center decouples
at lower temperature. These differences would make the spectra
steeper, but their effect is canceled by the edges of the system
freezing out at a higher temperature.

\begin{figure}[t]
\centering
\epsfig{file=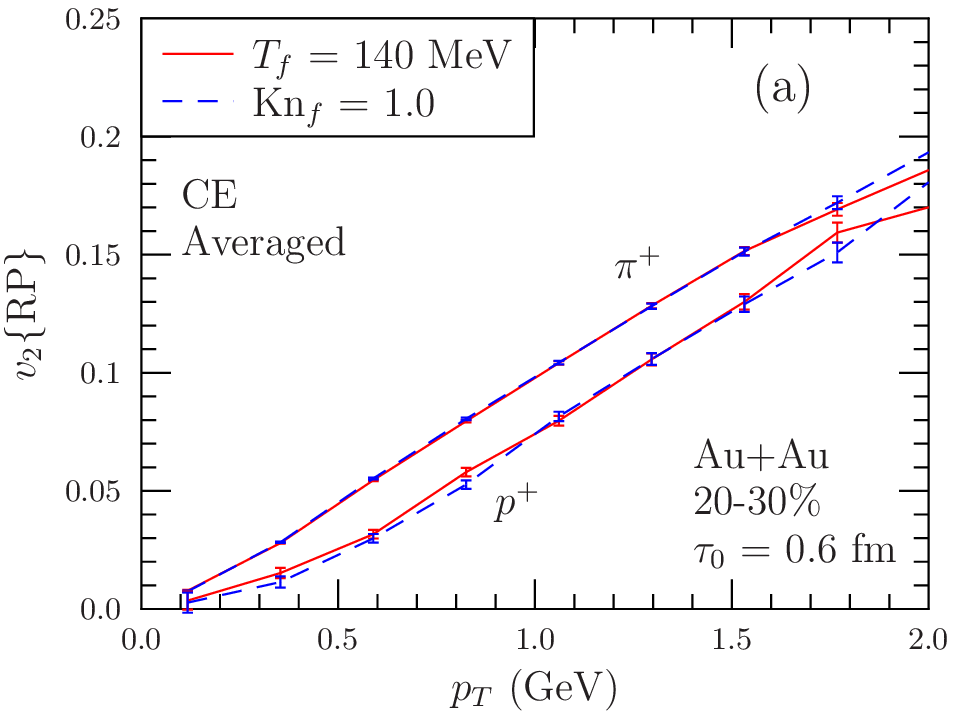, width=7.5cm}
\epsfig{file=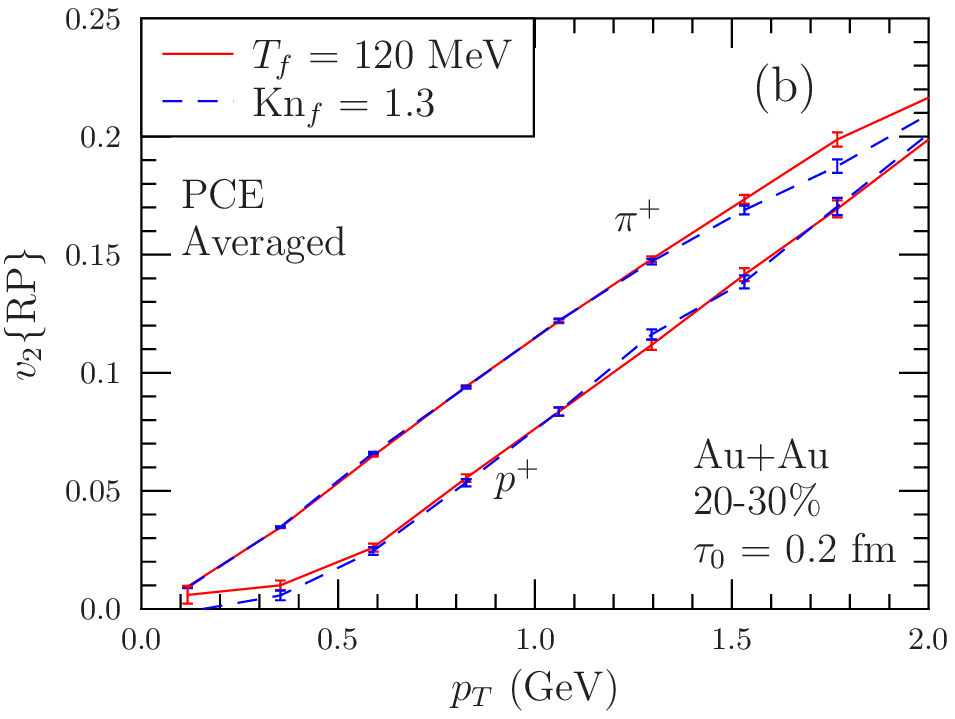, width=7.5cm}
\caption{Elliptic flow of positively charged pions and protons in
$\sqrt{s_{NN}}=200$~GeV 20-30\% central Au\,+\,Au collisions assuming
(a) chemical or (b) partial chemical equilibrium in the
EoS. The solid red line corresponds to the results obtained using
freeze-out at constant temperature and the dashed blue line using
freeze-out at constant Knudsen number. Error bars depict estimated
statistical errors.}
\label{fig: rhic v2}
\end{figure}

Next in Fig.~\ref{fig: rhic v2} we plot the $p_T$-differential
elliptic flow $v_2(p_T)$ of pions and protons at
$\sqrt{s_{NN}}=200$~GeV 20-30\% central Au\,+\,Au collisions using both
EoSs and freeze-out criteria. Since we use an averaged initial state,
we have evaluated the elliptic flow with respect to the reaction
plane, $v_2\{\text{RP}\}$.

In our earlier proceedings contribution~\cite{Holopainen:2013jna}, we
saw that elliptic flow was sensitive to the freeze-out criterion when
\emph{s95p}-PCE-v1 EoS was used. However, in that calculation we had
fixed $\Kn_f = 1.0$, and the $p_T$ distributions were different as
well. Now, after choosing the freeze-out Knudsen number to reproduce
the data and the spectra calculated using the constant temperature
freeze-out criterion, both freeze-out criteria lead to similar
elliptic flow. The same happens also when we keep $\Kn_f = 1.0$ fixed,
and adjust the freeze-out temperature instead to $T_f = 140$ MeV to
achieve similar spectra.

To study whether the sensitivity to the freeze-out criterion might
depend on the initial state, we performed the calculations using pure 
binary-collision profile as well. We used an initial time $\tau_0 = 0.6$~fm,
freeze-out temperature $T_f=120$~MeV, and Knudsen number $\Kn_f=1.3$
with \emph{s95p}-PCE-v1, and found that the spectra and elliptic
flow were again independent of the freeze-out criterion. Thus we
suspect that this similarity with both criteria is not due to some
property of the initial state but could be a more general
phenomenon. Also note that the same pair of constant temperature and
Knudsen number worked with both initial states.

\subsection{Event-by-event fluctuating initial states in 
$\sqrt{s_{NN}}=200$~GeV Au+Au collisions at RHIC}

\begin{figure}[b]
\centering
\epsfig{file=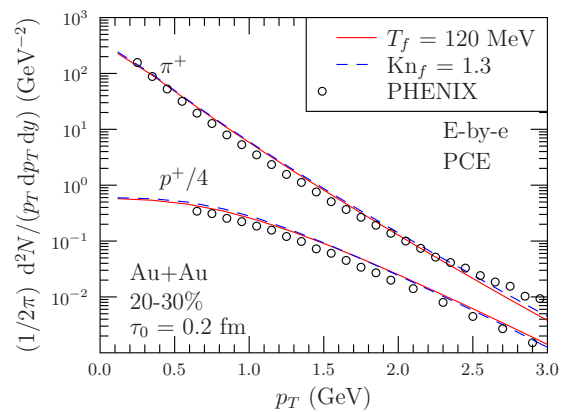, width=7.5cm}
\caption{Transverse momentum spectra of positively charged pions and
protons in $\sqrt{s_{NN}}=200$~GeV 20-30\% central Au\,+\,Au
collisions from event-by-event hydrodynamical simulations. The
solid red line corresponds to the results obtained using
freeze-out at constant temperature and the dashed blue line using
freeze-out at constant Knudsen number. The data are from the
PHENIX Collaboration \cite{Adler:2003cb}.}
\label{fig: rhic spectra ebye}
\end{figure}

As argued in the introduction, in event-by-event calculations the two
freeze-out criteria might lead to different results, even if the
results were similar when averaged initial state was used. To study
this assumption, we modeled the collisions at RHIC event-by-event
using the chemically frozen \emph{s95p}-PCE-v1 EoS. We followed the
same procedure than in our calculations using an averaged initial
state, and treated both the freeze-out temperature and Knudsen number
as free parameters to be adjusted to reproduce the observed $p_T$
spectra. It turned out that the same combination of parameters, 
$T_f = 140$ MeV and $\Kn_f = 1.3$, lead to a reasonable reproduction
of the data in both event-by-event and averaged initial state
calculations, see Figs.~\ref{fig: rhic spectra ebye} and~\ref{fig:
rhic spectra}, respectively. However, as observed before, e.g., in
Ref.~\cite{Holopainen:2010gz}, the spectra are a little bit flatter in
the event-by-event case.

\begin{figure}[t]
\centering
\epsfig{file=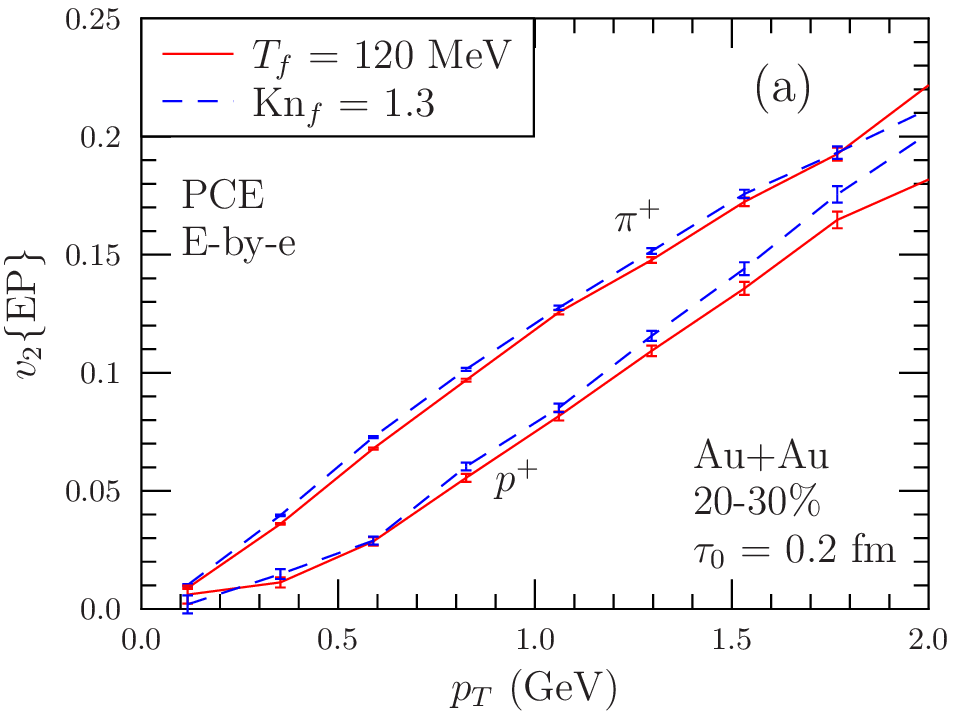, width=7.5cm}
\epsfig{file=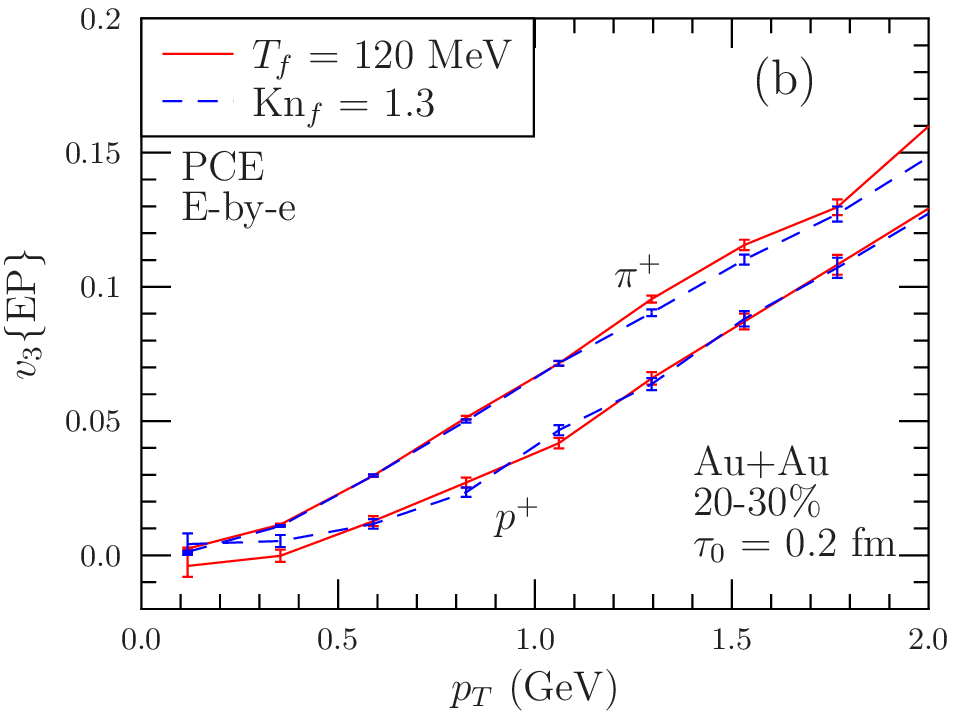, width=7.5cm}
\caption{(a) Elliptic and (b) triangular flow of positive
pions and protons in $\sqrt{s_{NN}}=200$~GeV 20-30\% central Au\,+\,Au
collisions from event-by-event hydrodynamical simulations. The
solid red line corresponds to the results obtained using
freeze-out at constant temperature and the dashed blue line using
freeze-out at constant Knudsen number. Error bars depict estimated
statistical errors.}
\label{fig: rhic vn ebye}
\end{figure}

The $p_T$-differential elliptic and triangular flows are shown in
Fig.~\ref{fig: rhic vn ebye}. In event-by-event calculations it makes
more sense to calculate the flow coefficients with respect to their
event planes, and therefore $v_n\{EP\}$ values are shown in the
figures. Consequently, comparison to the averaged initial state case
cannot be made, because the definitions of flow are different.

Unlike what we expected, there is no significant difference between
the freeze-out criteria. We also checked with a smaller number of events
that in the most-central collisions, where both $v_2$ and $v_3$ are
generated mostly by fluctuations, the situation is the same. Thus both
anisotropies seem to be insensitive to the freeze-out criterion in
event-by-event calculations too.

We have also checked that in individual events, the spectra, elliptic
flow, and triangular flow are not necessarily the same with the
parameters used, but the difference can be of the order of 10\% in
each studied variable. This opens up the question whether the
event-by-event distribution of anisotropies~\cite{Niemi:2012aj} might
be sensitive to the freeze-out criterion, and how the freeze-out
criterion would affect the correlation between the initial state
anisotropy and final momentum
anisotropy~\cite{Niemi:2012aj,Niemi:2015qia}. We have not checked
either what would happen if we adjusted the freeze-out criteria
event-by-event so that the $p_T$ distributions were similar in
each single event.

\subsection{Averaged initial state in $\sqrt{s_{NN}}=2760$~GeV Pb+Pb 
collisions at the LHC}

At a single collision energy one can always fix the freeze-out
temperature to reproduce the $p_T$ spectra, but there is no physical
reason why the same freeze-out temperature should work at another
collision energy. On the other hand, the dynamical criterion with
freeze-out at constant Knudsen number is based on local expansion
dynamics and general considerations about the validity of hydrodynamics,
and therefore we can expect the freeze-out to take place at the same
value of Knudsen number independent of the collision energy. Thus it
is worthwhile to check what happens in collisions at the LHC energy.

\begin{figure}[t]
\centering
\epsfig{file=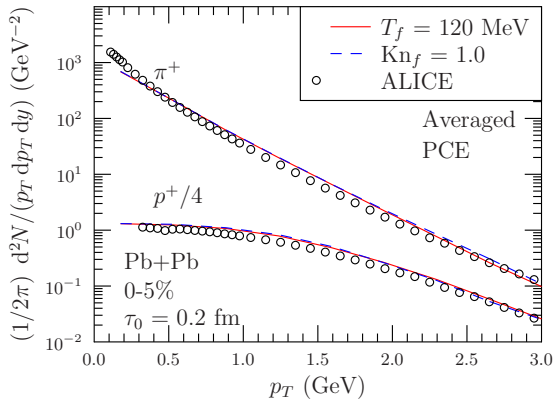, width=7.5cm}
\caption{Transverse momentum spectra of positive pions and protons
in $\sqrt{s_{NN}}=2760$~GeV 0-5\% central Pb\,+\,Pb collisions. The
solid red line corresponds to the results obtained using
freeze-out at constant temperature and the dashed blue line using
freeze-out at constant Knudsen number. The data are from the
ALICE Collaboration \cite{Abelev:2012wca}.}
\label{fig: lhc spectra smooth}
\end{figure}

In Fig.~\ref{fig: lhc spectra smooth} we show the transverse momentum
spectra of pions and protons in $\sqrt{s_{NN}}=2760$~GeV 0-5\% central
Pb\,+\,Pb collisions using averaged initial state and \emph{s95p}-PCE-v1
EoS. Both in the shown 0-5\% centrality class, and in the semicentral
20-30\% centrality class, the favored freeze-out temperature was the
same $T_f = 120$ MeV both at RHIC and at the LHC, but the data favored
lower freeze-out Knudsen number $\Kn_f = 1.0$ at the LHC. Thus, as
expected, the freeze-out Knudsen number does not depend on the
centrality of the collision, but contrary to expectations, it depends
on the collision energy. The dependence on collision energy may be an
effect of neglecting dissipation: When the dynamical criterion of
freeze-out at constant Knudsen number was used in the context of
dissipative hydro~\cite{Molnar:2014zha}, the same freeze-out Knudsen
number worked both at RHIC and the LHC. On the other hand, since the
slopes of the final $p_T$ distributions depend on the initial pressure
gradients, the collision energy dependence of the freeze-out Knudsen
number may also indicate that our Glauber-based initial-state model
does not reproduce the initial gradients properly. Thus it would be
interesting to apply the dynamical freeze-out criterion to more
sophisticated
EKRT~\cite{Niemi:2015qia,Eskola:1999fc,Paatelainen:2012at} and
IP-Glasma~\cite{ipglasma,ipglasma2} initial states.

\begin{figure}[t]
\centering
\epsfig{file=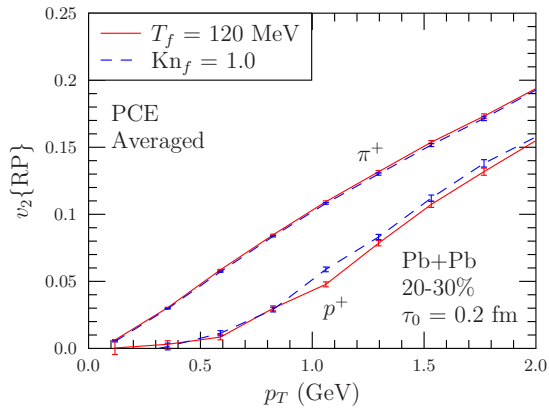, width=7.5cm}
\caption{Elliptic flow of positive pions and protons in
$\sqrt{s_{NN}}=2760$~GeV 20-30\% central Pb\,+\,Pb collisions. The
solid red line corresponds to the results obtained using
freeze-out at constant temperature and the dashed blue line using
freeze-out at constant Knudsen number. Error bars depict estimated
statistical errors.}
\label{fig: lhc v2 smooth}
\end{figure}

The $p_T$-differential elliptic flow of pions and protons shown in
Fig.~\ref{fig: lhc v2 smooth} depicts the same pattern at the LHC as
at RHIC: Once the $p_T$ spectra are reproduced, both freeze-out
criteria lead to similar elliptic flow. 

To be sure, we carried out the event-by-event calculations at the LHC
energy too, but saw the very same behavior as at RHIC and when using
the averaged initial state: Once the freeze-out parameters were chosen
to reproduce the observed spectra ($T_f = 120$ MeV and $\Kn_f = 1.0$),
the elliptic and triangular flows were similar too.

\section{Conclusions}

As argued in the introduction, freeze-out criterion based on
freeze-out at a constant temperature is an oversimplification, and a
dynamical criterion where freeze-out takes place at constant Knudsen
number would be more physical. However, we saw that in semicentral
and central collisions, identified particle spectra and elliptic and
triangular flows are not sensitive to the freeze-out criterion.

We evaluated the Knudsen number as the ratio of the expansion rate of
the system, and the scattering rate of pions. We applied the
freeze-outs at constant temperature and constant Knudsen number to
ideal fluid calculations of Au\,+\,Au collisions at RHIC and Pb\,+\,Pb
collisions at the LHC at 20-30\% centrality, and fixed the values of
freeze-out parameters by fitting the observed pion and proton $p_T$
distributions. The two criteria lead to different freeze-out surfaces:
with dynamical freeze-out the edges decouple earlier (\emph{i.e.},~at
higher temperature) and the center of the system lives longer, letting
the matter cool more compared to the constant temperature
case. However, after the $p_T$ spectra were constrained to be similar,
no sign of the different temperatures and flow velocities on the
freeze-out surface could be seen in the anisotropies.

We did check that the same insensitivity persists in most-central
collisions, but we did not check what might happen when the collision
system is much smaller such as in peripheral $A+A$ or in $p+A$
collisions. The earlier results of Ref.~\cite{Molnar:2014zha} indicate
that the sensitivity to the freeze-out criterion increases when the
system size or collision energy decreases, and thus the $p+A$ collision
system could be very sensitive to the freeze-out criterion. Maybe even
to such an extent that the Knudsen number at the very beginning of the
evolution is larger than one~\cite{Niemi:2014wta}.

Our event-by-event calculations revealed that even if the spectra and
anisotropies after averaging over many events were not sensitive to
the freeze-out criterion, spectra and anisotropies in individual
events were. This leaves open the question whether event-by-event
distributions of average $p_T$ or anisotropy coefficients $v_n$ might
be sensitive to the freeze-out criterion. One could also expect that
HBT radii would be an observable which is more sensitive than the
anisotropies to the exact properties of the freeze-out surface.

Unfortunately we were unable to study how the value of the smearing
parameter $\sigma$ of the Monte Carlo Glauber model affects the
sensitivity to freeze-out, and thus whether small scale density
fluctuations in the initial state might affect the freeze-out. This
remains to be explored in a further study, although one may expect
that dissipation has largely smeared away small scale structures by
the time of freeze-out.

\newpage

\begin{acknowledgments}
Fruitful discussions with S.~Bass, M.~Bleicher, H.~van~Hees,
H.~Honkanen, J.~I.~Kapusta, A.~Kurkela, D.~Molnar, B.~Tomasik, and
R.~Venugopalan are gratefully acknowledged. We thank J.~Jankowski,
M.~A.~R.~Kaltenborn, M.~Marczenko, and H.~Niemi for careful reading of
the manuscript and constructive comments. PH.~thanks for hospitality
the Iowa State University and Goethe University where part of this
work was done. During the long gestation of this project, the work of
PH.~has been supported by the National Science Center, Poland, under
grant Polonez DEC-2015/19/P/ST2/03333 funded from the European
Union's Horizon 2020 research and innovation programme under the Marie
Sk\l odowska-Curie grant agreement No 665778, the National Science
Center, Poland, under Maestro grant DEC-2013/10/A/ST2/00106, BMBF
under Contract No.~06FY9092, the ExtreMe Matter Institute (EMMI), DOE
Grant DE-AC02-98CH10886, and Johannes R\aa ttendahl foundation. The
works of SA.~and HH.~have been supported by DOE Grant DE-FG02-01ER41200
and the ExtreMe Matter Institute (EMMI), respectively.
\end{acknowledgments}

\appendix
\section{Integrals in the calculation of the scattering rate}
\label{algebra}

The reduction of the number of integrals over momentum in the
scattering-rate calculations has been shown in
Ref.~\cite{Zhang:1998tj} for equal-mass particles obeying Boltzmann
statistics, and generalized for nonidentical particles when the
scattering partner has a fixed momentum in
Ref.~\cite{Tomasik:2002qt}. For the sake of completeness, we repeat
the process here, and generalize it for quantum statistics.

The total number of times pions scatter with particles $i$ per unit
volume per unit time is given by:
\begin{equation}
\begin{split}
{\cal R}_i =&\, 2\int_{s_a}^\infty 
\dif s\,\sqrt{(s-s_a)(s-s_b)}\,\sigma_{\pi i}(s)\\ 
&\times\int\frac{\dif^3 p_\pi}{2E_\pi} \frac{\dif^3 p_i}{2E_i}\, 
f_\pi(T,\mu_\pi) f_i(T,\mu_i)\, 
\delta\!\left(s - (p_i + p_\pi)^2\right),
\end{split}
\end{equation}
where, compared to Eq.~(\ref{rate1}), we have
added the integration over center-of-mass energy $s$ and the
corresponding $\delta$ function. To proceed we express the
distribution functions $f_\pi$ and $f_i$ as series:
\begin{equation}
\begin{split}
f_i(T,\mu_i)& = \frac{g_i}{(2\pi)^3} \frac{1}{e^{\frac{E-\mu_i}{T}}\pm 1}\\
& = \frac{g_i}{(2\pi)^3}
\sum_{n=1}^\infty (\mp 1)^{n+1} e^{n\mu_i/T} e^{-nE/T},
\end{split}
\end{equation}
where $-1$ in the series is for fermions and $+1$ for bosons,
change the momentum coordinates to spherical coordinates, change the
integral over the magnitude of momentum to integral over energy, and
rewrite the $\delta$-function as
\begin{equation}
\begin{split}
&\delta\!\left(s - (p_i + p_\pi)^2\right)\\ 
& = \frac{1}{2|\mathbf{p}_\pi||\mathbf{p}_i|}
  \delta\!\left(\cos \theta_i + \frac{s-(m_\pi^2+m_i^2)-2E_\pi E_i}
{2|\mathbf{p}_\pi||\mathbf{p}_i|} \right).
\end{split}
\end{equation}
The angular integrals can now be carried out, and we get
\begin{equation}
\begin{split}
{\cal R}_i = \frac{g_i}{2^5\pi^4}&\sum_{k=1}^\infty e^{k\mu_\pi/T}
\sum_{n=1}^\infty (\mp 1)^{n+1} e^{n\mu_i/T}\\
&\times\int_{s_a}^\infty \dif s\,\sqrt{(s-s_a)(s-s_b)}\,\sigma_{\pi i}(s)\\
&\times\int_{m_\pi}^\infty\dif E_\pi \int_{m_i}^\infty \dif E_i\,
e^{-\frac{k}{T}\left(E_\pi + \frac{n}{k}E_i\right)}\\
&\times\Theta\!\left(1-\left|\frac{s-(m_\pi^2+m_i^2)-2E_\pi E_i}
{2|\mathbf{p}_\pi||\mathbf{p}_i|}
\right|\right).
\end{split}
\end{equation}
We change the integration variables from $E_\pi$ and $E_i$ to 
$y = E_\pi + \frac{1}{r}E_i$ and $x = E_\pi - \frac{1}{r}E_i$, where $r = k/n$.
The $\Theta$ function constraint can now be written as $b<x<c$, where
\begin{equation}
\begin{split}
b = & \frac{(r^2 m_\pi^2 - m_i^2)y - d}{rs-(r-1)(m_i^2-rm_\pi^2)}\\
c = & \frac{(r^2 m_\pi^2 - m_i^2)y + d}{rs-(r-1)(m_i^2-rm_\pi^2)}\\
d = & \sqrt{r^2y^2-(rs-(r-1)(m_i^2-rm_\pi^2))}\\
&\times\sqrt{(s-s_a)(s-s_b)}.
\end{split}
\end{equation}
It turns out that the integration over $x$ is constrained more by the
$\Theta$ function than by the integration limits, and we get
\begin{equation}
\label{step6}
\begin{split}
{\cal R}_i = \frac{g_i}{2^6\pi^4}&\sum_{k=1}^\infty e^{k\mu_\pi/T}
\sum_{n=1}^\infty (\mp 1)^{n+1} e^{n\mu_i/T}\\
&\times r\int_{s_a}^\infty \dif s\,\sqrt{(s-s_a)(s-s_b)}\,
\sigma_{\pi i}(s)\\
&\times\int_\alpha^\infty\dif y\, e^{-\frac{ky}{T}} (c-b),
\end{split}
\end{equation}
where
\begin{equation}
\alpha = \sqrt{\frac{s}{r}-\frac{r-1}{r^2}\left(m_i^2-rm_\pi^2\right)}.
\end{equation}
The $y$ integral can now be reordered and carried out to be
\begin{equation}
\label{K1}
\int_\alpha^\infty \dif y\, e^{-k\frac{y}{T}} \sqrt{y^2-\alpha^2} 
= \frac{T\alpha}{k}\, K_1\!\left(\frac{k\alpha}{T}\right),
\end{equation}
where $K_1$ is the modified Bessel function. Inserting this into 
Eq.~(\ref{step6}) and keeping the $y$-independent terms omitted from 
Eq.~(\ref{K1}), we finally get
\begin{equation}
\begin{split}
{\cal R}_i = \frac{g_i T}{2^5\pi^4}&\sum_{k=1}^\infty e^{k\mu_\pi/T}
\sum_{n=1}^\infty \frac{(\mp 1)^{n+1}}{n} e^{n\mu_i/T}\\
&\times \int_{s_a}^\infty \dif s\,\frac{(s-s_a)(s-s_b)\,\sigma_{\pi i}(s)}
{\sqrt{rs-(r-1)(m_i^2-rm_\pi^2)}} \\
&\times K_1\!\left(\frac{n}{T}\sqrt{rs-(r-1)(m_i^2 - rm_\pi^2)}\right).
\end{split}
\end{equation}
After summing over all particles $i$, and dividing by pion density, we
get Eq.~(\ref{rate}).

\section{Center-of-mass momentum in particle-resonance scattering}
\label{pcms}

If one of the scattering partners is a resonance, the conventional
expression for the center-of-mass momentum of the scattering,
\begin{equation}
\label{eq:pcms}
\begin{split}
p_\cms&\left(\sqrt{s},m_1,m_2\right)\\
& = \frac{\sqrt{(s-(m_1+m_2)^2)(s-(m_1-m_2)^2)}}{2\sqrt{s}},
\end{split}
\end{equation}
must be amended to take into account the finite width of
the resonance. To do this, we again mostly follow the UrQMD
description~\cite{Bass:1998ca}, and include an integral over the mass
distribution of the resonance:
\begin{equation}
\begin{split}
p_\cms\left(\sqrt{s}\right) = \int_0^{\sqrt{s}-m_\pi}\!\!\!\!\!&\dif m\,\, 
p_{CMS}\left(\sqrt{s},m_\pi,m\right)\\
&\times\frac{1}{2\pi}\frac{\Gamma_R}{(m_R-m)^2+\Gamma_R^2/4},
\end{split} 
\end{equation}
where we assume the mass distribution to be the Breit-Wigner
distribution with mass-independent width $\Gamma_R$, and $m_R$ is the
pole mass of the resonance.

Note that, in the integrals of Appendix~\ref{algebra}, and in the
evaluation of the EoS, the resonances have been assumed to have zero
width, and their pole masses have been used as their masses.

\section{Full decay width}
\label{gamma}

The evaluation of the full decay width $\Gamma_{\text{tot}}(M)$ in
Eq.~(\ref{cross-section}) requires knowledge of partial decay widths of
three- and four-body decay channels as well. Unfortunately
Eq.~(\ref{width}) cannot be easily generalized to many-body decays.
To treat all decay channels in a similar fashion, we combine the
particles in three- and four-body decays into a particle and a
particle pair, or two particle pairs, respectively, use the invariant
mass(es) of particle pair(s) to evaluate the center-of-mass momentum
(Eq.~(\ref{eq:pcms})), and use the available phase space to give the
mass distribution of the invariant mass of the pair(s). In particular,
for three-body decays we obtain
\begin{equation}
\begin{split}
p_\cms(M) = \frac{8M}{N} \int_{m_1+m_2}^{M-m_3}\!\!\!\!\!\dif m_{pair}&\,
[p_\cms(M,m_{pair},m_3)]^2\\
&\!\!\!\!\times p_\cms(m_{pair},m_1,m_2),
\end{split}
\end{equation}
where the normalization factor $N$ is given by
\begin{equation}
\begin{split}
N = 8M\int_{m_1+m_2}^{M-m_3}\!\!\!\!\!\dif m_{pair}&\,
p_\cms(M,m_{pair},m_3)\\
&\!\!\!\!\!\!\times p_\cms(m_{pair},m_1,m_2).
\end{split}
\end{equation}
If any of the daughter particles in a multiparticle decay is a
resonance, we use its pole mass only and neglect its width.

\end{document}